# Primordial Nucleosynthesis in Conformal Weyl Gravity

Lloyd Knox[*] and Arthur Kosowsky[†]

*Department of Physics*
*Enrico Fermi Institute, The University of Chicago, Chicago, IL 60637-1433*
*and*
*NASA/Fermilab Astrophysics Center*
*Fermi National Accelerator Laboratory, Batavia IL 60510-0500*
(October, 1993)

## Abstract

Recently conformal Weyl gravity has been considered as a candidate alternative gravity theory. This fourth-order theory is attractive because it is the only metric theory of gravity which is invariant under local conformal transformations of the metric. We calculate the primordial light element abundances in this theory. The major difference from the standard cosmology is that the universe expands far more slowly throughout the nucleosynthesis epoch. The production of $^4$He depends strongly on $\eta$, the ratio of baryons to photons. For $\eta = 10^{-8}$ the mass fraction of $^4$He is $X_4 \simeq 0.25$ and the number densities relative to hydrogen for $^2$H, $^3$He and $^7$Li are $n(^2\text{H})/n(\text{H}) \simeq 9 \times 10^{-20}$, $n(^3\text{He})/n(\text{H}) \simeq 4 \times 10^{-18}$ and $n(^7\text{Li})/n(\text{H}) \simeq 10^{-13}$. This value of $\eta$ corresponds to a baryon mass density close to the standard model critical density. However, adjusting $\eta$ to give a reasonable helium yield forces the deuterium and lithium yields to be small enough that the theory cannot be reconciled with observations.

04.50.+h, 98.80.Ft

Typeset using REVTEX

---

[*]knox@oddjob.uchicago.edu

[†]NASA GSRP Fellow. arthur@oddjob.uchicago.edu

## I. INTRODUCTION

Any alternative gravity theory must face the constraints imposed by the abundances of light elements, in addition to the classical solar system tests. Astronomical observations indicate that $^4$He, $^2$H, $^7$Li, and possibly $^3$He were produced at some time prior to the first stellar processing of nuclei. The standard model of cosmology explains these primordial abundances as remnants of the hot early stage of the universe, produced in the first few minutes after the bang (see, *e.g.*, [1,2]). Theory and observation are in concordance for the abundances of all four of these elements, abundances which span nine orders of magnitude [3]. For an alternative theory of gravity to be viable it must also accomodate these observations. Here we calculate the light element abundances produced in "conformal cosmology", the homogeneous and isotropic cosmology associated with conformal Weyl gravity [4]. Although the motivation for this article is the particular theory of conformal Weyl gravity, our results apply generally to any model resulting in a very slowly expanding universe.

Conformal Weyl gravity [5,6] is a metric theory invariant under conformal transformations of the metric, $g_{\mu\nu}(x) \to \Omega^2(x) g_{\mu\nu}(x)$. It follows from the Weyl action:

$$I_W = -\alpha \int d^4 x (-g)^{1/2} C_{\lambda\mu\nu\kappa} C^{\lambda\mu\nu\kappa} \qquad (1)$$

where $C_{\lambda\mu\nu\kappa}$ is the Weyl tensor, defined as

$$C_{\lambda\mu\nu\kappa} = R_{\lambda\mu\nu\kappa} - \frac{1}{2}(g_{\lambda[\nu} R_{\kappa]\mu} - g_{\mu[\nu} R_{\kappa]\lambda}) + \frac{1}{3} R g_{\lambda[\nu} g_{\kappa]\mu}. \qquad (2)$$

Note that this theory of gravity differs from general relativity only in the relation between curvature and the stress tensor: the Einstein equations are replaced by analagous fourth-order equations. All vacuum solutions of general relativity are also vacuum solutions of conformal Weyl gravity, although the latter theory also possesses additional vacuum solutions having no general relativistic counterparts. Mannheim and Kazanas [7] have found spherically symmetric, static solutions to conformal Weyl gravity which correspond to the Schwarzschild solution at small distances. On larger scales the theory deviates from general relativity; it is claimed that this deviation can account for flat galactic rotation curves [8].

Here we are interested in the isotropic and homogeneous solutions to these equations. As in general relativity, the most general solutions are the Robertson-Walker spacetimes. However, the conformally invariant theory possesses only the trivial vacuum solution $T_{\mu\nu} = 0$. Mannheim [9] introduces a scalar field conformally coupled to gravity and to the matter fields which spontaneously breaks the conformal invariance; its non-zero vacuum expectation value is $S$. The resulting expansion rate is given by

$$H^2 = \left(\frac{\dot{R}}{R}\right)^2 = -\frac{2\rho}{S^2} - \frac{k}{R^2} - 2\lambda S^2. \qquad (3)$$

where $\lambda$ is the quartic self-coupling of the scalar field and $k = -1$, an open universe, since the right side of Eq. (3) must be positive. Note that the energy density contributes negatively to the expansion rate, a general feature of fourth-order gravity theories. Mannheim [4] considers this to be an advantage because it provides a resolution to the flatness problem.

The scalar field expectation value $S$ is a free parameter; however, we do not need to know its value to calculate the results of primordial nucleosynthesis [10]. Inspection of Eq. (1) shows that the Universe has a minimum size and it is always curvature dominated [11] Taking the minimum size to be well before nucleosynthesis, and assuming that the $\lambda$ term is not significant at the present epoch, the expansion rate from the epoch of nucleosynthesis to the present time is well approximated by $H^2 = 1/R^2$. Thus in the absence of any significant entropy production,

$$H = H_0(R_0/R) = H_0(T/T_0) = 1.1 \times 10^{-9} h T_9 \text{ sec}^{-1}. \tag{4}$$

A zero subscript denotes the value of a quantity today, h is the Hubble constant today in units of 100 km/sec/Mpc, and $T_9$ is the temperature of the Universe in units of $10^9$ K. For comparison, in the standard model, $H \simeq 10^{-2} T_9^2 \text{ sec}^{-1}$. We have assumed $\lambda$ can be neglected. In this model the deceleration parameter is $q_0 = \lambda/H_0^2$ and hence the value of $\lambda$ is constrained by kinematic tests such as the galaxy number count vs. redshift test. Taking a conservative limit of $q_0 < 3$, we find $1/R_0^2 < 4H_0^2$ and hence the expansion rate during nucleosynthesis can be no more than twice as fast as in the $\lambda = 0$ case. A factor of two increase in the expansion rate will have no qualitative effects on our results.

## II. NUCLEOSYNTHESIS IN CONFORMAL WEYL GRAVITY

To calculate the abundances of light elements produced in a universe described by conformal Weyl gravity, we modified the expansion rate in the nucleosynthesis code of Kawano [12]. The solutions presented here used a reaction network of 18 nucleides. All of the following calculations take a Hubble constant of $h = 1$. The resulting mass fraction of $^4$He and the number densities relative to hydrogen for the various nucleides as a function of $\eta$ are displayed in Fig. 1. In the following discussion, $X_2$, $X_t$, $X_3$ and $X_4$ refer to the mass fractions of $^2$H, $^3$H, $^3$He and $^4$He respectively. As in the standard case, the only nucleide produced in substantial quantity is $^4$He; its abundance increases with increasing baryon to photon ratio. Note that for $\eta \simeq 10^{-8}$, the helium mass fraction falls in the observed range, although the deuterium mass fraction is many orders of magnitude below its standard value for any $\eta$. Before discussing implications, we give a qualitative explanation of these results.

It may seem surprising that the observed helium mass fraction can be produced with an expansion rate that is ten million times slower than in the standard case. It is well known that relatively minor changes in the expansion rate — for example, due to an extra neutrino species — significantly alter the $^4$He predictions. However, it is easy to understand why there is another slower value of the expansion rate for which $X_4 \simeq 1/4$. In the standard case, the $^4$He production occurs after the Universe has already left nuclear statistical equilibrium (NSE). The first reaction rates to become slow compared to the expansion rate (causing a departure from NSE) are the weak rates which interconvert protons and neutrons. This occurs at a temperature of about $T_9 = 10$. At about $T_9 = 3$ the species favored by NSE changes from protons to $^4$He but significant $^4$He production must wait until $T_9 \approx 1$. At this temperature the $^2$H abundance becomes large enough to fuel the $^4$He production; virtually all the neutrons that exist when $^4$He synthesis commences are turned into $^4$He. As the expansion rate is slowed from that of the standard case, the $^4$He yield is lowered because the neutrons and protons stay in equilibrium down to a lower temperature, reducing the neutron abundance. However, this decrease in the $^4$He yield for slower expansions cannot continue indefinitely. If the Universe expands slowly enough, the universe will track equilibrium into the era in which $^4$He is favored by equilibrium and everything will turn into $^4$He. Thus it makes sense that there exists a significantly slower expansion rate that gives the appropriate $^4$He abundance.

To understand the results in more detail we consider the evolution of the various abundances with respect to their NSE values. We expect NSE to obtain when the rates for all relevant reactions are faster than the expansion rate. The mass fractions in NSE are given by [1]

$$X_j = \left(\frac{g_j}{2}\right) \left(\frac{\zeta(3)2^{3/2}T^{3/2}\eta}{\sqrt{\pi}m^{3/2}}\right)^{A_j-1} A_j^{5/2} X_p^{Z_j} X_n^{A_j-Z_j} \exp(B_j/T), \tag{5a}$$

$$X_n = X_p \exp(-Q/T), \tag{5b}$$

$$1 = X_p + X_n + \sum_i X_i, \tag{5c}$$

where $B_j$, $A_j$, $Z_j$, and $g_j$ are the binding energy, atomic weight, atomic number, and spin multiplicity of nucleide $j$, $m$ is the nucleon mass, and $Q$ is the neutron-proton mass difference. Comparison of the NSE abundances and the actual abundances calculated by evolving the rate equations reveals the physical reasons for the final abundances.

A good approximation to NSE for our purposes is to truncate the sum in Eq. (5c) at $^4$He and only consider protons, neutrons, deuterium, $^3$H, $^3$He and $^4$He. Assuming NSE abundances for all the elements we can then calculate reaction rates. The only process that significantly depletes $^4$He is photodestruction, $\gamma + {}^4\text{He} \leftrightarrow \text{n} + {}^3\text{He}$. The rate for this process becomes slower than the expansion rate at $T_9 = 4.4$; at this temperature, the universe first leaves NSE. Figure (2) shows that the $X_4$ fails to keep pace with the increasing equilibrium value for temperatures $T_9 < 4.4$. The production of $^4$He continues out of equilibrium until the weak rates that interconvert protons and neutrons freeze out at $T_9 \simeq 1.6$. The value of $X_4$ at $T_9 = 4.4$ varies as $\eta^3$ as expected from Eq. (5a). This power law for the final abundances is modulated from $\eta^3$ to approximately $\eta^2$ because $X_3$ stops tracking its equilibrium value for $T_9 \lesssim 4$.

In the cosmology of Weyl conformal gravity, far less deuterium is produced than in the standard case. This is due to two factors. First, the proton-neutron ratio tracks NSE to much lower temperatures in the conformal case, so when synthesis of deuterium and helium starts to be significant, far fewer neutrons are available to be converted to deuterium. In effect, all available neutrons go into helium via deuterium, but compared to the standard case, this process is slow. The weak reaction rates continually replenish the small neutron NSE abundance; all available neutrons dribble into helium, but at any given temperature the amount of deuterium is small compared to the standard case. Second, once the dominant deuterium production process $\text{p} + \text{n} \leftrightarrow {}^2\text{H} + \gamma$ freezes out at $T_9 \simeq 1.0$, the residual amount of deuterium continues decreasing by the process $^2\text{H} + {}^3\text{He} \rightarrow \text{p} + {}^4\text{He}$ which does not freeze out until $T_9 \simeq 0.15$. This process remains effective for a much longer time than in the standard scenario, resulting in a much larger suppression of $X_2$. From a heuristic viewpoint, a universe with a very slow expansion rate suppresses the amount of neutrons when nucleosynthesis begins; a helium mass fraction of around 0.25 can be built up slowly, but then the residual mass fractions of the lighter nucleides, particularly deuterium, are much smaller than in the standard case.

## III. IMPLICATIONS AND CONCLUSION

The primordial abundance of $^4$He is estimated from observations of extragalactic HII regions. The signature of $^4$He in these regions is its recombination radiation, and thus neutral helium is invisible. The amount of neutral helium must be modeled, which necessarily introduces systematic errors. It is difficult to determine the uncertainty in these measurements, but various estimates are all in agreement. As a representative value, the review by Walker et al. [3] takes $X_4 = 0.23 \pm 0.01$. Conformal Weyl gravity can indeed produce this amount of helium for $\eta \simeq 10^{-8}$.

Despite this modest and somewhat surprising success, the inability to produce significant deuterium is a major problem for the theory. Since deuterium is so weakly bound it is difficult to create in significant amounts in any astrophysical process other than the big bang. Because of this, observations of deuterium are believed to provide firm lower limits to the primordial value of $n(^2H)/n(H)$ [13,14]. The best measurements of $n(^2H)/n(H)$ come from Hubble Space Telescope data. Linsky et al. [15] have determined $n(^2H)/n(H) = 1.65^{+0.07}_{-0.18} \times 10^{-5}$ in our local interstellar medium along a particular line of sight. Due to destruction of deuterium during the evolution of the galaxy this is probably 1.5 to 3 times smaller than the primordial value. However, for our puposes factors of two are irrelevant; $n(^2H)/n(H)$ as calculated here misses the mark by over 14 orders of magnitude.

Production of $^7$Li also presents a problem, though not as glaring as the deuterium discrepancy. $^7$Li is both produced and destroyed in stars. By plotting $n(^7Li)/n(H)$ vs. metallicity one sees the "Lithium plateau" at low metallicity, suggesting a primordial abundance of $n(^7Li)/n(H) \simeq (8 \pm 1) \times 10^{-11}$ derived as an average from 35 measurements of old stars [3]. The value from conformal Weyl gravity falls short of this mark by about three orders of magnitude for $\eta = 10^{-8}$.

These results do not categorically rule out a theory of gravity described by Eq. (1). It is conceivable that, given a primordial abundance of $^4$He, some as yet undiscovered process could produce $^2$H and $^7$Li prior to stellar nucleosynthesis. For example, Dimopoulos et al. [16] have invoked a late decaying particle to catalyze nuclear reactions, resulting in deuterium production. However, the simplicity and success of the standard cosmology strongly disfavors such poorly motivated speculations. We also note that in order for conformal Weyl gravity to have a non-trivial cosmology, the conformal invariance must be broken. We have considered the particular model advanced by Mannheim, in which a conformally coupled scalar field spontaneously breaks the symmetry. In the context of current understanding of particle theories, this procedure is the simplest and most natural. More complicated mechanisms of symmetry breaking may lead to an equation for the scale factor evolution with different properties than Eq. (3). Again, however, the fact that simple general relativity leads naturally to successful cosmological models disfavors this possibility.

We have shown that, as expected, a universe which expands very slowly compared to the standard scenario can produce a primordial abundance of $^4$He consistent with current observations. However, the appropriate amount of helium in a slowly expanding universe necessarily implies a vastly underproduced deuterium abundance, as well as a lithium shortage. This result casts serious doubt on the viability of conformal Weyl gravity. In addition, our results strongly support the claim that the universe must expand at a rate very near that given by the standard cosmology to pass the primordial nucleosynthesis test.

## ACKNOWLEDGMENTS

We thank Geza Gyuk and Brian Fields for their assistance with the nucleosynthesis code. AK is supported by the NASA Graduate Student Research Program. This work has been partially supported by the DOE and by NASA grant NAGW 2381 at Fermilab.

FIGURES

Mass fraction of $^4$He and number densities relative to hydrogen for $^3$He, $^7$Li and $^2$H as functions of $\eta$.

Numerical solution of nucleide mass fractions as a function of temperature, along with NSE abundance of $^4$He (dashed line), for $\eta = 10^{-8}$.

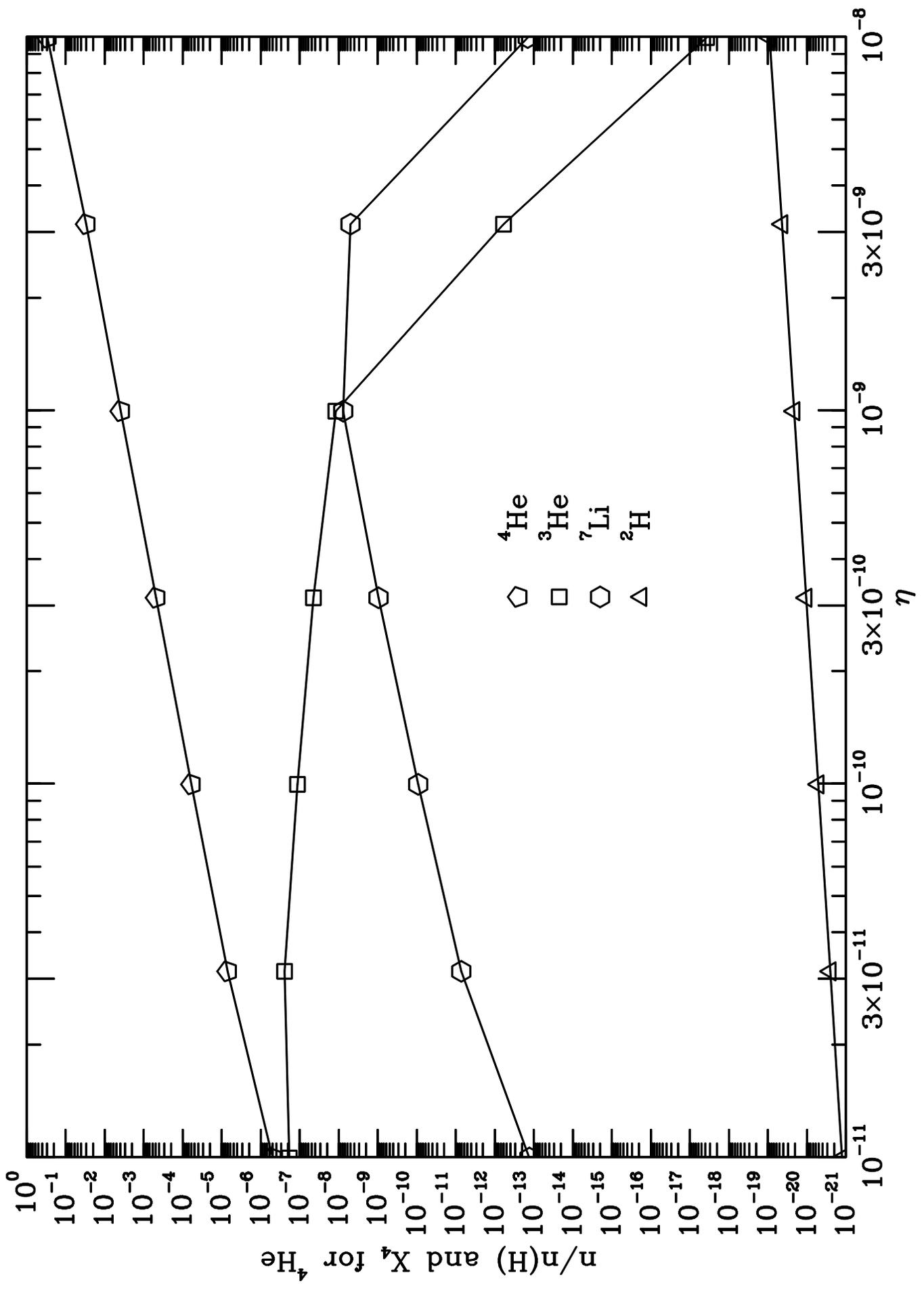

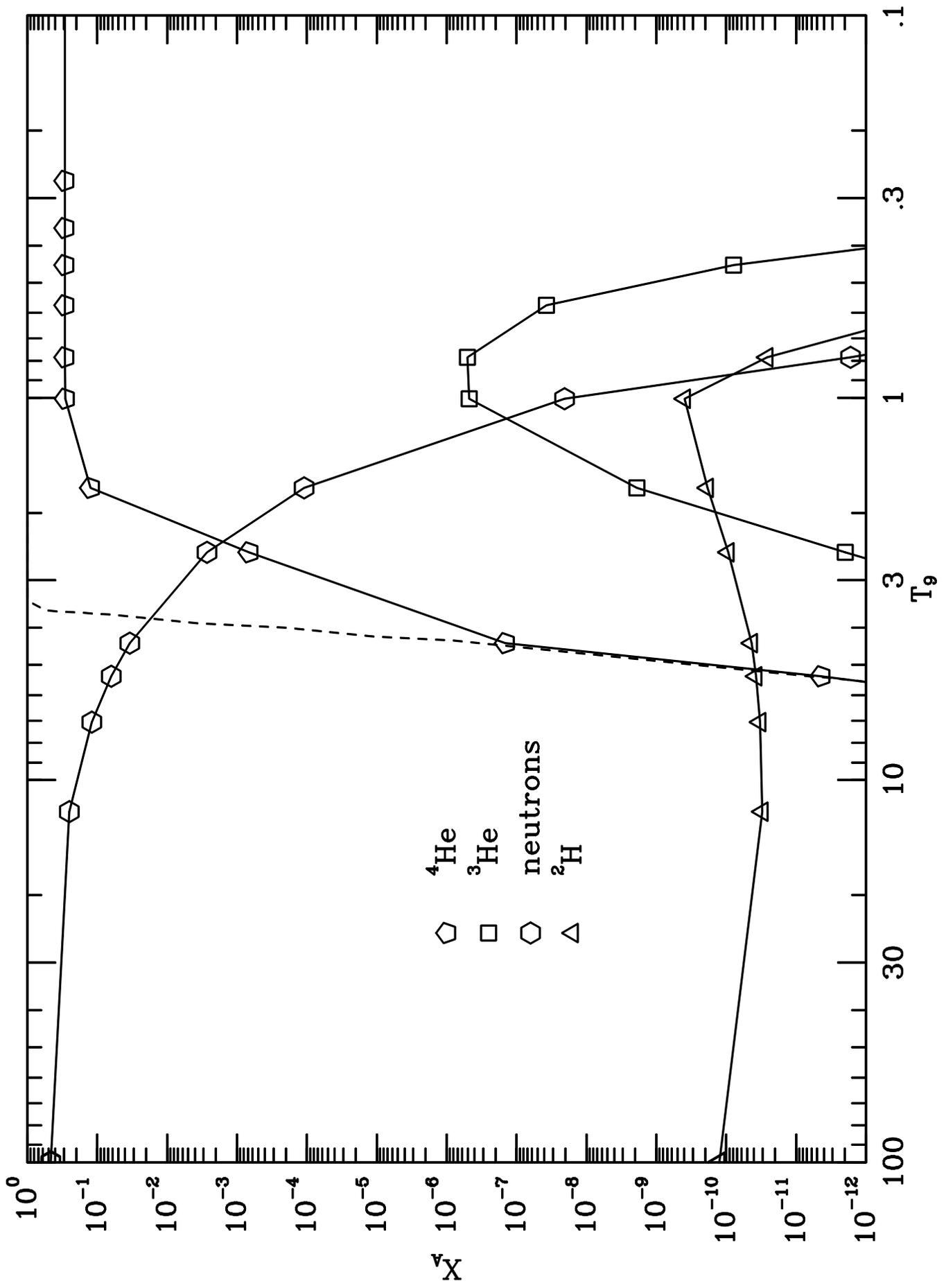